\documentclass[letterpaper]{article} %
\usepackage{aaai2026}  %
\usepackage{verbatim}
\usepackage{times}  %
\usepackage{helvet}  %
\usepackage{courier}  %
\usepackage[hyphens]{url}  %
\usepackage{graphicx} %
\usepackage{natbib}  %
\usepackage{caption} %
\usepackage{algorithm}
\usepackage{algorithmic}
\usepackage{amssymb}
\usepackage{booktabs}
\usepackage{tabularx}
\usepackage{longtable}
\usepackage{listings}
\usepackage{xcolor}
\usepackage{array}
\usepackage{multirow}
\usepackage{caption}
\usepackage{fancyvrb}
\usepackage{mdframed}
\usepackage{tcolorbox}
\usepackage{float}

\definecolor{lightgray}{RGB}{245,245,245}
\definecolor{variablecolor}{RGB}{0.2,0.9,0.9}
\definecolor{highlightcolor}{RGB}{0.8,0,0}
\usepackage{newfloat}
\usepackage{listings}
\usepackage{amsmath}

\lstdefinestyle{verilog}{
    language=Verilog,
    basicstyle=\ttfamily\footnotesize,
    keywordstyle=\color{blue}\bfseries,
    commentstyle=\color{green!60!black},
    stringstyle=\color{red},
    backgroundcolor=\color{gray!10},
    frame=single,
    breaklines=true,
    breakatwhitespace=false,
    tabsize=4,
    showstringspaces=false,
    numbers=left,
    numberstyle=\tiny\color{gray},
    captionpos=b
}

\usepackage{siunitx}

\usepackage{booktabs}
\DeclareCaptionStyle{ruled}{labelfont=normalfont,labelsep=colon,strut=off} %
\floatstyle{ruled}
\newfloat{listing}{tb}{lst}{}
\floatname{listing}{Listing}
\usepackage{xcolor}
\usepackage{colortbl}
\newcounter{corrfn}
\def\correspondingauthor{%
      \ifnum\value{corrfn}=0%
        \footnote{Corresponding author.}%
        \setcounter{corrfn}{\value{footnote}}%
      \else%
        \footnotemark[\value{corrfn}]%
      \fi%
    }%
\title{Principle-Guided Verilog Optimization: IP-Safe Knowledge Transfer via Local-Cloud Collaboration}
\author{
    Jing Wang\textsuperscript{\rm 1,2}\equalcontrib,
    Zheng Li\textsuperscript{\rm 2}\equalcontrib,
    Lei Li\textsuperscript{\rm 3},
    Fan He\textsuperscript{\rm 2},\\
    Liyu Lin\textsuperscript{\rm 2},
    Yao Lai\textsuperscript{\rm 3},
    Yan Li\textsuperscript{\rm 2}\correspondingauthor,
    Xiaoyang Zeng\textsuperscript{\rm 2},
    Yufeng Guo\textsuperscript{\rm 1}
}
\affiliations{
    \textsuperscript{\rm 1}Nanjing University of Posts and Telecommunications, 
    \textsuperscript{\rm 2}Fudan University\\
    \textsuperscript{\rm 3}The University of Hong Kong\\

    \texttt{wangjing25@njupt.edu.cn \quad 24212020117@m.fudan.edu.cn \quad liyan@fudan.edu.cn}
}

\usepackage{bibentry}

\begin{document}

\maketitle

\begin{abstract}
Recent years have witnessed growing interest in adopting large language models (LLMs) for Register Transfer Level (RTL) code optimization. While powerful cloud-based LLMs offer superior optimization capabilities, they pose unacceptable intellectual property (IP) leakage risks when processing proprietary hardware designs. In this paper, we propose a new scenario where Verilog code must be optimized for specific attributes without leaking sensitive IP information.
We introduce the first IP-preserving edge-cloud collaborative framework that leverages the benefits of both paradigms. Our approach employs local small LLMs (e.g., Qwen-2.5-Coder-7B) to perform secure comparative analysis between paired high-quality target designs and novice draft codes, yielding general design principles that summarize key insights for improvements. These principles are then used to query stronger cloud LLMs (e.g., Deepseek-V3) for targeted code improvement, ensuring that only abstracted and IP-safe guidance reaches external services.
Our experimental results demonstrate that the framework achieves significantly higher optimization success rates compared to baseline methods. For example, combining Qwen-2.5-Coder-7B and Deepseek-V3 achieves a 66.67\% optimization success rate for power utilization, outperforming Deepseek-V3 alone (49.81\%) and even commercial models like GPT-4o (55.81\%). Further investigation of local and cloud LLM combinations reveals that different model pairings exhibit varying strengths for specific optimization objectives, with interesting trends emerging when varying the number of comparative code pairs.
Our work establishes a new paradigm for secure hardware design optimization that balances performance gains with IP protection.
Our code and dataset are available at \url{https://github.com/friyawang/VeriOptim}.

\end{abstract}

\section{Introduction}

\begin{figure}[t!]
    \centering
    \includegraphics[width=\columnwidth]{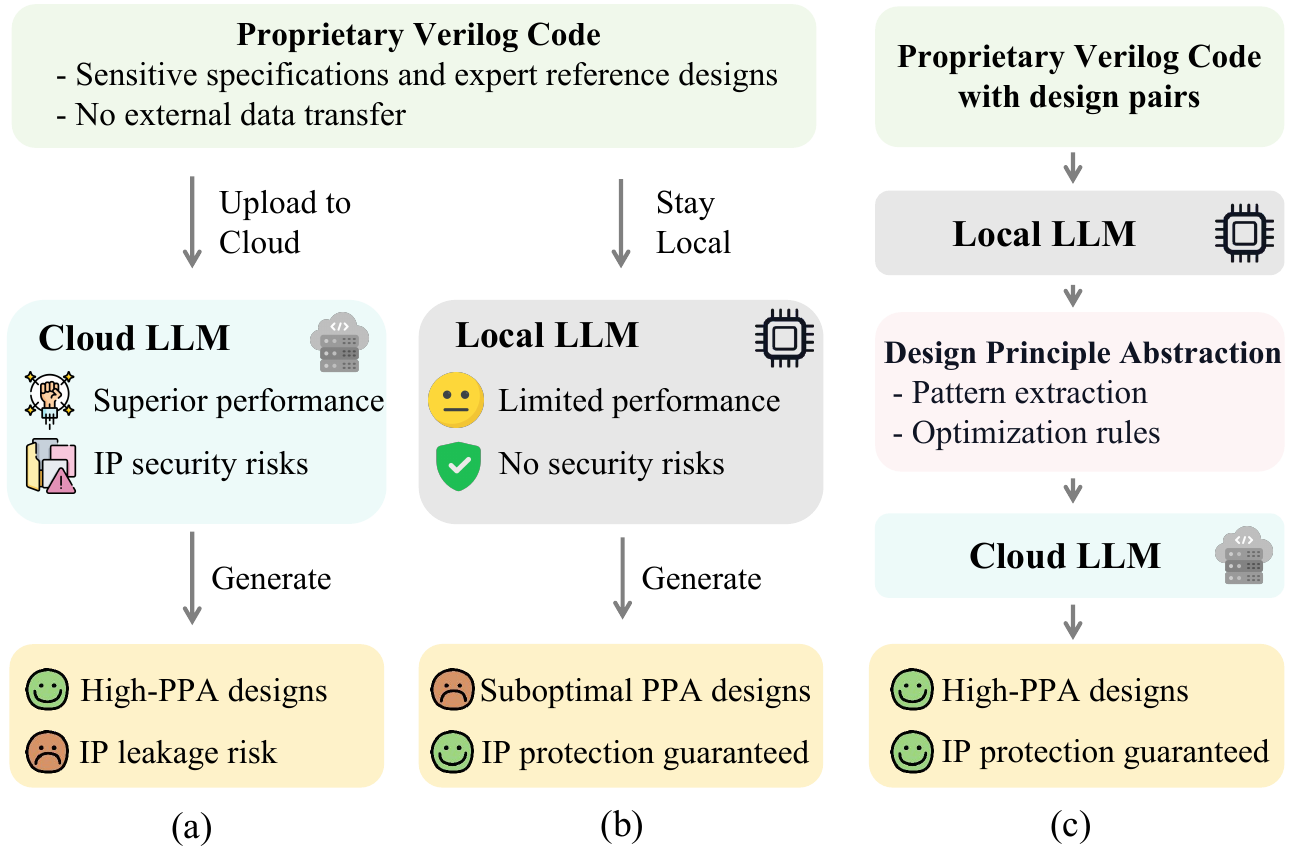}
    \caption{Comparison of LLM Deployment Strategies: Traditional approaches (a) Direct cloud processing with IP Risk, (b) Local-only processing with performance constraints, and (c) Our proposed Local-Cloud framework optimizes the hardware design without leaking IP.}
    \label{fig:llm_deployment}
\end{figure}

Large Language Models (LLMs)~\citep{radford2019gpt,gemini,qwen2} are bringing changes to Electronic Design Automation (EDA)~\citep{Wu2023ChatEDAAL,Zhong2023LLM4EDA,Lai2024AnalogCoderAC}, particularly in digital IC design workflows. These models have demonstrated capabilities in generating and optimizing Hardware Description Language (HDL) code, effectively producing Register Transfer Level (RTL) code from functional specifications~\cite{RTLCoder,li2024rtllm,Pei2024BetterV}. This technological advancement has gained traction in industrial settings, where LLMs increasingly serve as assistive tools for engineers in code development and design optimization tasks.

The effectiveness of LLM-based optimization in digital chip design hinges critically on the quality of contextual examples provided to the model. Design optimization typically targets specific attributes—such as power consumption or critical path delay (where $f_{max} = 1/t_{cp}$, with $f_{max}$ representing maximum clock frequency and $t_{cp}$ denoting critical path delay)—which inherently involve complex trade-offs. Proprietary IP code that embodies proven optimization strategies for these objectives would constitute an invaluable context, as such examples encapsulate years of accumulated engineering expertise and validated design patterns.

However, a fundamental tension emerges between leveraging proprietary knowledge and maintaining intellectual property (IP) security. In the semiconductor industry, IP represents a critical strategic asset underpinning competitive advantage~\cite{rajendran2017overview,zamiri2023ip,knechtel2019protect}. State-of-the-art LLMs operate predominantly as cloud-based services, introducing substantial security vulnerabilities when handling sensitive design data, as evidenced by incidents of inadvertent proprietary code exposure~\cite{kuteyi2023samsung}. Consequently, semiconductor companies have implemented stringent policies restricting proprietary code transmission to external platforms~\cite{chui2023state}.
While locally deployed LLMs address privacy concerns, they face significant performance limitations due to constrained computational resources compared to cloud-based counterparts~\cite{isachenko2024generative}. As illustrated in Figure~\ref{fig:llm_deployment}, this binary choice between high-performance cloud solutions with security risks and secure local deployments with limited capabilities represents a significant barrier to widespread LLM adoption in enterprise EDA environments. Existing research~\cite{thoratllm,tasnia2025veriopt,chipdesignagent} mainly focuses on model effectiveness, while inadequately addressing this critical trade-off between privacy protection and performance.

To bridge this divide, we propose a novel IP-preserving edge-cloud collaborative framework for Verilog optimization that leverages principle extraction techniques to synergistically combine the advantages of local and cloud-based LLMs. 
Specifically, our framework operates through a two-stage edge-cloud collaboration: (1) a local LLM (e.g., Qwen2.5-Instruct-7B~\citep{Yang2024Qwen25}) extracts generalizable design principles by analyzing proprietary-draft code pairs on-premise, ensuring sensitive IP never leaves the secure environment, and (2) a powerful cloud LLM (e.g., Deepseek-V3~\citep{DeepSeekV3TR}) applies these abstracted principles to optimize new designs without accessing any proprietary implementations. This design strategically combines the advantages of both model types: smaller local models provide secure access to proprietary knowledge for principle extraction, while large cloud models contribute their superior code generation capabilities to apply these principles effectively. The principle extraction mechanism enables the transfer of knowledge from high-quality proprietary code while maintaining complete IP isolation from external cloud services.

To enable comprehensive evaluation of hardware optimization methods, we curate a novel contrastive PPA dataset comprising 1,196 power-optimized and 967 timing-optimized Verilog implementation pairs across nine design categories. Our dataset addresses key limitations in existing benchmarks by providing functionally equivalent design variants with standardized Performance-Power-Area metrics extracted through systematic synthesis validation. 
Using this curated dataset, we conduct a comprehensive evaluation of our IP-preserving optimization framework. Our experimental evaluation demonstrates that the framework achieves substantial optimization success rates of 50.85\% for critical path delay and 66.67\% for power optimization, representing significant improvements over standalone cloud models (33.90\% and 49.81\% respectively) and surpassing proprietary baselines like GPT-4o. 
Notably, our method enables weaker open-source models to outperform stronger proprietary alternatives while maintaining complete IP security. Case studies reveal dramatic improvements, including 22.83\% critical path delay reduction and up to 81.77\% power savings through comprehensive architectural restructuring guided by extracted principles. However, our analysis also uncovers critical limitations: while the framework excels at local, pattern-based optimizations (e.g., 84.8\% success on regular structures like counters), it struggles with global architectural changes and functional semantics, with success rates dropping significantly for timing-critical tasks requiring deep hardware understanding. These findings highlight both the promise and current boundaries of LLM-based hardware optimization, revealing interesting model-specific sensitivities to the number of in-context examples that inform optimal deployment strategies.

In summary, our contributions are two-fold: (1) \textbf{We formulate a novel IP-preserving Verilog optimization task} that addresses the critical industrial challenge of leveraging powerful cloud-based LLMs while maintaining complete intellectual property security. (2) \textbf{We develop a comprehensive solution framework} comprising a curated contrastive PPA dataset and an edge-cloud collaborative optimization method that strategically combines local principle extraction with cloud-based code generation, achieving substantial improvements and revealing both the capabilities and limitations of our framework.
\begin{figure*}[!t]
    \centering
    \includegraphics[width=0.9\textwidth]{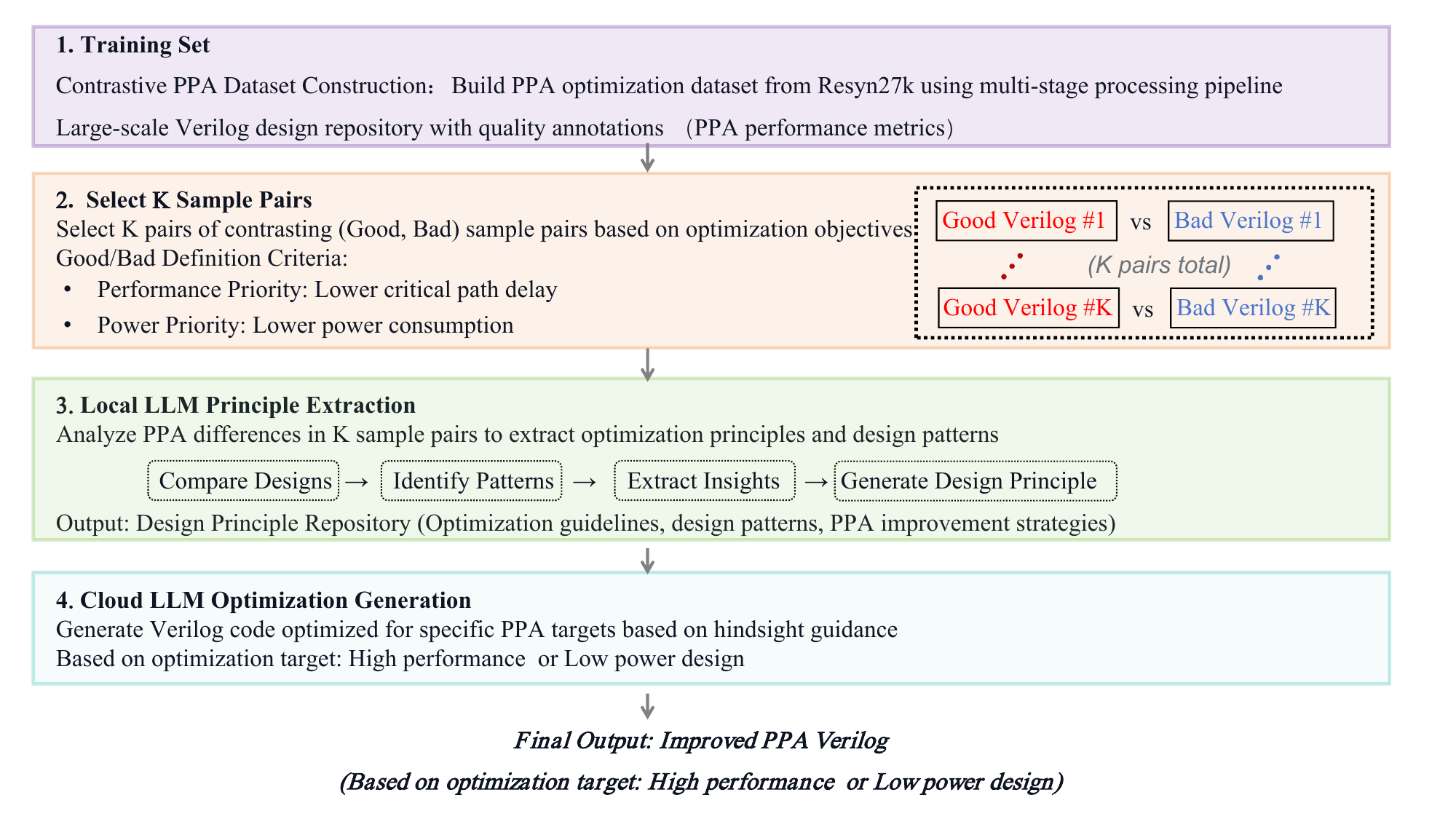}
    \caption{Principle Extraction-Based Code Optimization Pipeline Using Contrastive PPA Analysis}
    \label{fig:hindsight_optimization}
\end{figure*}

\section{Related Work}
The integration of LLMs into EDA has emerged as a promising paradigm for revolutionizing hardware design workflows. Recent research efforts have primarily focused on two complementary streams. The first centers on \emph{dataset curation and benchmarking}, where researchers have developed specialized corpora such as VeriGen \cite{thakur2024verigen}, comprehensive evaluation benchmarks like RTLLM \cite{li2024rtllm} spanning from combinational logic to complex state machines, and RLTRepo~\citep{Allam2024RTLRepo} for large-scale RTL design projects. Data augmentation techniques have also been explored, with \citet{chang2024data} proposing automated frameworks incorporating Verilog abstract syntax tree enhancement to improve model training.
The second stream encompasses \emph{AI/LLM-aided frameworks for diverse EDA tasks}, including specialized code generation models such as RTLCoder \cite{RTLCoder}, which outperformed GPT-4 on VerilogEval benchmarks, and MEV-LLM \cite{Nadimi2024}, integrating multiple fine-tuned models for targeted learning. Beyond code generation, these frameworks extend to optimization through tools like RTL Rewriter \cite{RTLwriter} employing retrieval-augmented generation, debugging solutions including RTLFixer~\citep{tsai2024rtlfixer} and MEIC~\citep{xu2024meic}, verification automation via LLM4DV~\citep{zhang2025llm4dv} and AutoSVA~\citep{orenes2021autosva}, and physical implementation tasks, demonstrating the breadth of AI applications across the EDA workflow.

However, the highly sensitive nature of intellectual property in the semiconductor industry creates a fundamental barrier: companies cannot share proprietary design data with cloud-based LLMs due to security concerns, limiting the practical deployment of these powerful optimization tools in real-world scenarios. In this paper, we explore this more practical setting and propose a local-cloud collaborative framework that protects IP while leveraging powerful cloud models to improve hardware design optimization.

\section{Method}

We address the challenge of optimizing Verilog code for hardware attributes while protecting intellectual property. 
The key to our framework is edge-cloud collaboration, where local LLMs extract generalizable design principles from proprietary code, and cloud LLMs apply these principles to optimize new designs. Figure~\ref{fig:hindsight_optimization} illustrates the complete pipeline.

\subsection{Problem Definition}
Consider two codebases: a proprietary codebase $\mathcal{P} = \{p_1, p_2, \ldots, p_K\}$ containing optimized Verilog modules with sensitive IP, and a draft codebase $\mathcal{D} = \{d_1, d_2, \ldots, d_K\}$ containing functionally equivalent but lower-quality implementations that can be publicly shared. Each pair $(p_i, d_i)$ represents the same functionality, where $p_i$ demonstrates superior performance in target attributes (e.g., power consumption, critical path delay).
The core challenge is: \textit{How can we leverage the optimization insights from proprietary-draft pairs to improve new Verilog designs without exposing the proprietary code to external systems, e.g., commercial LLMs such as GPT-4o?}

\subsection{Task Formulation}

Given an attribute evaluation function $f: \mathcal{V} \rightarrow \mathbb{R}^+$ that measures performance (higher is better), we have $f(p_i) > f(d_i)$ for all pairs. For a new Verilog module $n$ requiring optimization, our goal is to generate an improved version $n^*$ such that $f(n^*) > f(n)$, subject to the constraint that the proprietary codebase $\mathcal{P}$ must never be accessible to cloud services.
The key insight is leveraging the performance gaps between proprietary and draft implementations to guide optimization. 
This creates a fundamental challenge: we need to transfer optimization knowledge from $\mathcal{P}$ while ensuring complete IP protection.

\begin{table*}[t!]
    \centering
    \small 
    \begin{tabular}{@{}l|cc|cc@{}}
    \toprule

     \textbf{Model} & \textbf{CPD SR (\%)}  & \textbf{Rel. Improv. (\%)} & \textbf{Power SR (\%)} & \textbf{Rel. Improv. (\%)} \\
     \midrule
      \rowcolor{gray!15} GPT-4o (Direct)  & 45.76 &27.31 &  55.81&36.44 \\ 
     \rowcolor{gray!15} Claude-3.7-Sonnet (Direct)&52.54 & 29.67& 67.79 &43.17  \\ 
     \rowcolor{gray!15}  Gemini-2.5-Pro (Direct)  &57.63 & 68.81& 66.67 &46.32  \\ 
          \midrule
     
     Qwen2.5-Instruct-7B  (w / IP data)& 40.68  & 36.30 & 34.83  &  30.91\\ 
     Qwen2.5-Coder-7B (w / IP data)& 38.98 & 47.48 & 36.70&  9.50\\%
     DeepSeek-V3 (w / IP data)& 37.29 &27.98& \underline{65.92}&32.23\\ 
     DeepSeek-R1 (w / IP data) &  \underline{50.85} &29.72 &63.30&27.55\\ 
     \midrule 
      DeepSeek-V3 (Direct) & 33.90 &25.10& 49.81&22.81\\ 

     Ours (Coder + V3) & 44.07 &31.30 & 66.67 & 34.42\\ 
     Ours (Instruct + V3) & \underline{\textbf{50.85}}&  31.03 & \underline{\textbf{66.67}}& \textbf{35.29}\\ 
     \midrule 
     DeepSeek-R1 (Direct) &  40.68&33.90 & 48.31& \textbf{36.86}\\ 
     
     Ours (Coder + R1) & \underline{\textbf{44.07}}& 28.98 & \underline{\textbf{57.30}} &28.43\\ 
     Ours (Instruct + R1) & 44.07 & 27.80 & 55.06& 30.23\\ 
     \bottomrule
    \end{tabular}
    \caption{Power and critical path delay optimization success rate results using different local models and a combination of cloud-edge models. Our method achieves a significant gain over using local models without leaking sensitive IP, even outperforming the cloud LLMs with access to IP data. Results of commercial models are reference only (in gray). CPD: Critical Path Delay.}
    \label{tab:main_ret}
\end{table*}

\subsection{Local-Cloud Collaborative Framework}

We propose a two-stage approach that resolves this challenge through principle abstraction, leveraging the unique capabilities of local and cloud LLMs:

\paragraph{Stage 1: Local Principle Extraction}
A local LLM (e.g., Qwen2.5-7B-Instruct~\citep{Yang2024Qwen25}) is deployed on-premise with full access to the proprietary codebase. These small models can run entirely on local hardware without any external communication, ensuring that sensitive IP never leaves the secure environment. The local LLM analyzes $K$ proprietary-draft pairs to extract design principles:
\begin{equation}
H = \text{LocalLLM}(\{(p_i, d_i)\}_{i=1}^K, P_1) = \{h_1, h_2, \ldots, h_M\}
\end{equation}
where $P_1$ is the prompt instructing the local LLM to compare proprietary and draft implementations, identify optimization patterns, and summarize them as abstract principles without revealing specific implementation details.
These principles $H$ capture optimization insights while satisfying two critical properties:
\begin{itemize}
\item \textbf{IP-safety:} Principles contain no proprietary code snippets or implementation details;
\item \textbf{Actionability:} Principles provide concrete guidance for improving draft designs.
\end{itemize}
For example, when optimizing for critical path delay consumption, a principle might be: \emph{Avoid using deeply nested logic in `assign` statements when possible}, rather than exposing specific proprietary implementations.
Note that the selection strategy of these $K$ pairs could influence the quality of the summarized design principles. 
In our current implementation, we employ a random selection for simplicity and investigate the effect of varying $K$ on optimization performance in the experimental section.
\paragraph{Stage 2: Cloud-Based Code Generation}
The extracted principles and the new module are sent to a powerful cloud LLM for optimization. These cloud models (e.g., Deepseek-V3 of 671B~\citep{DeepSeekV3TR} and GPT-4o) are typically hosted on external servers, offering superior code generation capabilities but posing IP risks if given direct access to proprietary code. By receiving only abstracted principles rather than actual proprietary implementations, the cloud LLM can leverage its advanced capabilities to generate optimized code:
\begin{equation}
n^* = \text{CloudLLM}(n, H, P_2)
\end{equation}
where $P_2$ is the prompt instructing the cloud LLM to apply the extracted principles $H$ to optimize the target module $n$ for specific hardware attributes (e.g., power consumption or critical path delay).
This separation ensures that while we benefit from state-of-the-art cloud models, the proprietary codebase remains completely isolated from external services. The optimized module $n^*$ is subsequently synthesized and evaluated using hardware synthesis tools, with its performance metrics compared against the original module $n$ to determine whether $f(n^*) > f(n)$, thus validating the success of the optimization.

\section{Experiments}

\subsection{Dataset Curation}
To enable comprehensive evaluation, we curate a novel contrastive PPA dataset that addresses key limitations in existing benchmarks: limited scale, structural inconsistency, and absence of functionally equivalent design variants for direct performance comparison.

\paragraph{Dataset Construction Pipeline.} Our construction process follows a four-stage pipeline. Starting from the Resyn27k dataset~\citep{RTLCoder}, we first filter 8,531 syntactically correct and synthesizable Verilog modules using automated synthesis validation through Xilinx Vivado. Second, we classify samples by functional categories (multipliers, FIFOs, ALUs, etc.) and bit-width configurations to ensure functional equivalence within groups. Third, we identify optimization pairs using TF-IDF-based similarity scoring with a 0.7 threshold, ensuring contrastive samples differ primarily in optimization characteristics rather than functionality. Finally, we validate semantic consistency through manual inspection of representative subsets.
\paragraph{PPA Metric Extraction.} Each design undergoes standardized synthesis using Synopsys Design Compiler under uniform constraints (100 MHz target frequency, 28 nm technology node). We extract three key metrics: cell area ($\mu$m²), dynamic power consumption (mW), and critical path delay (ns). The final dataset comprises 1,196 power-optimized and 967 timing-optimized implementation pairs across nine design categories, where each data sample contains an instruction for the module, two Verilog implementations, and the corresponding PPA metrics.
Visualized samples can be found in the Appendix.

\subsection{Experimental Settings}

\paragraph{Metrics}
We evaluate the effectiveness of Verilog optimization using two complementary metrics.
\textbf{(i) Success Rate (SR)} measures the proportion of examples where optimization achieves improvement in the target metrics. 
We focus on the critical path delay and power consumption of the hardware design, as these two metrics are the most critical optimization targets in modern PPA analysis, with timing performance directly determining system clock frequency and power consumption being paramount for energy-efficient and thermally sustainable designs.
For both power consumption and critical path delay, success is defined as achieving lower values compared to the original implementation. The success rate is calculated as the ratio of successful optimizations to the total number of evaluated examples.
\textbf{(ii) Relative Improvement (Rel. Improv.)} quantifies the magnitude of optimization for successful cases, computed as:
\begin{equation*}
    \text{Relative Improv} = \frac{\text{Original} - \text{Optimized}}{\text{Original}} \times 100\%.
\end{equation*}
Both metrics are computed separately for power consumption and critical path delay to assess optimization effectiveness from different hardware design perspectives.

\begin{figure}[t!]
    \centering
    \includegraphics[width=\linewidth]{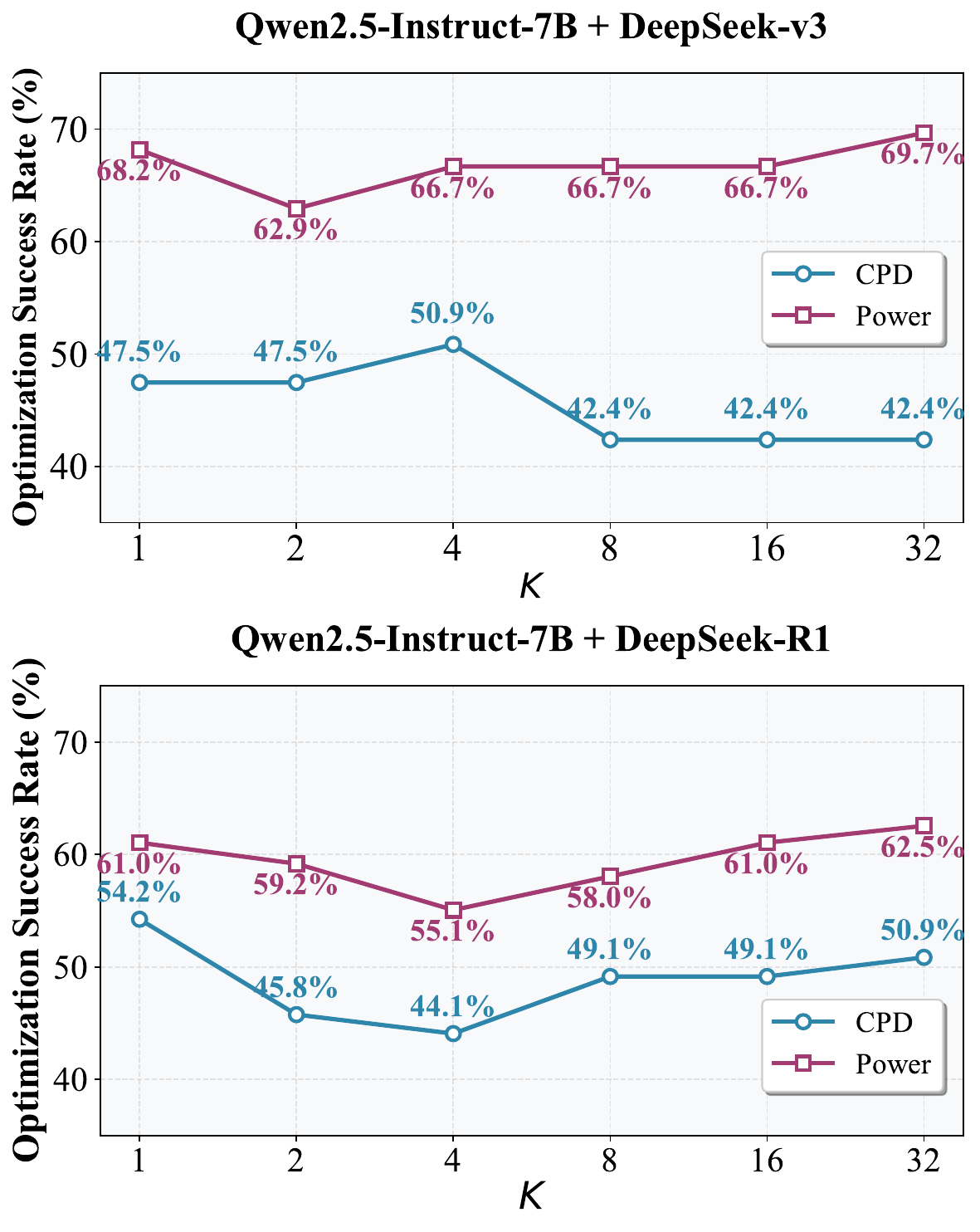}
    \caption{Effect of the number of in-context learning comparative pairs. }
    \label{fig:effect_varyK}
\end{figure}

\paragraph{Evaluated Models}
The models utilized in this work represent cutting-edge advances in language model architectures and training methodologies.
\textbf{Local LLMs:} (i) Qwen2.5-7B-Instruct~\citep{Yang2024Qwen25} is a 7.61 billion parameter instruction-tuned model that excels in language understanding, multilingual capabilities, coding, mathematics, and reasoning, supporting up to 128K token context length and over 29 languages, including Chinese and English.
(ii) Qwen2.5-7B-Coder~\citep{hui2024qwen25coder} is the specialized coding version of Qwen2.5, bringing significant improvements in code generation, code reasoning, and code fixing compared to its predecessor. 
\textbf{Cloud LLMs:} (i) DeepSeek-V3~\citep{DeepSeekV3TR} is a powerful Mixture-of-Experts (MoE) language model with 671 billion total parameters and 37 billion activated for each token, pre-trained on 14.8 trillion diverse and high-quality tokens, adopting Multi-head Latent Attention (MLA) and DeepSeekMoE architectures while pioneering an auxiliary-loss-free strategy for load balancing and multi-token prediction training objectives. (ii) DeepSeek-R1~\citep{guo2025deepseek} represents the state-of-the-art open-source reasoning model that achieves performance comparable to OpenAI-o1 across reasoning tasks, trained via large-scale reinforcement learning based on the base version of DeepSeek-V3.
The prompts used to extract design principles for local LLMs and to improve the Verilog code according to the summarized key insights for cloud LLMs can be found in the Appendix.
\paragraph{Compared Methods}
We compare our approach against three types of baseline methods. \emph{(1) Small LLMs:} We evaluate Qwen2.5-Instruct-7B and Qwen2.5-Coder-7B, which are directly prompted with IP-sensitive data for code optimization. These models can access sensitive information but have limited optimization capabilities. \emph{(2) Large LLMs:} We test large open-source models (DeepSeek-V3 and DeepSeek-R1) under two settings: (i) Direct, without access to sensitive IP data, where models are prompted directly to optimize the code, and (ii) with the same IP data used in our method (included for reference only, as this would violate IP confidentiality in practice). \emph{(3) Proprietary Models:} We evaluate leading commercial models including GPT-4o~\citep{gpt4o}, Claude-3.7-Sonnet~\citep{claude37extendthink}, and Gemini-2.5-Pro~\citep{gemini25}, which are directly prompted to optimize the Verilog code. Since the underlying training corpus and architecture design of these models remain unclear, their performance is for reference only.

\subsection{Main Results}

\begin{table}[t!]
    \centering
    \small
    \begin{tabular}{@{}l|cc|c@{}}
    \toprule
    \textbf{Case} & \textbf{Ori. PPA} & \textbf{Improved PPA} & \textbf{Imp. (\%)} \\
    \midrule
    \multicolumn{4}{c}{\textbf{Critical Path Delay Improvement}} \\
    \midrule
    adder\_4bit\_carry   & 320.00 ps & 310.00 ps & 3.12\% \\
    barrel\_shifter    & 290.00 ps & 270.00 ps & 6.90\% \\
    calculator         & 2.76 ns     & 2.13 ns     & 22.83\% \\
    \midrule
    \multicolumn{4}{c}{\textbf{Power Improvement}} \\
    \midrule
    division\_module    & 22.53 \textmu W & 21.15 \textmu W & 6.13\% \\
    schmitt\_trigger    & 1.66 \textmu W  & 0.89 \textmu W  & 46.39\% \\
    display\_transmitter & 70.49 \textmu W & 12.85 \textmu W & 81.77\% \\
    \bottomrule
    \end{tabular}
    \caption{PPA metrics improvement (critical path delay and power utilization) in the optimized RTL design.}
    \label{tab:case_study_ppa_improvement}
\end{table}

\begin{figure*}[t!]
    \centering
\includegraphics[width=\linewidth]{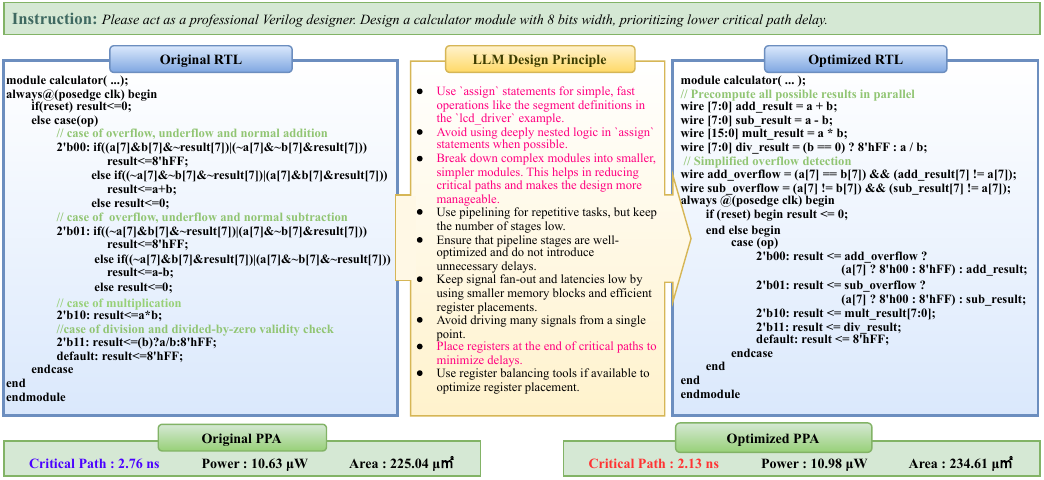}
    \caption{Optimization example of critical path optimization of our framework, where Qwen2.5-Instruct-7B summarizes key principles for optimization, successfully hinting Deepseek-V3 to reduce the critical path delay from 2.70 ns to 2.13 ns. The adopted principles are shown in pink.}
    \label{fig:case_study}
\end{figure*}

The main results of different methods are presented in Table~\ref{tab:main_ret}. We highlight the following key findings:
(i) \textbf{Local small models exhibit limited effectiveness in Verilog code optimization} despite having access to sensitive IP data. Qwen2.5-Instruct-7B and Qwen2.5-Coder-7B achieve optimization success rates of only around 35-40\% for both timing and power optimization. 
In contrast, \textbf{cloud-based models with IP data access demonstrate substantially superior performance}, e.g., DeepSeek-v3 achieving a 65.92\% power optimization success rate and R1 achieving a 50.85\% timing optimization success rate.
However, this approach is impractical for real-world deployment due to the significant security risks associated with uploading proprietary IP data to cloud servers, which motivates our exploration of secure alternatives that can leverage these powerful models.
(ii) \textbf{Our proposed method successfully bridges this gap by combining the security advantages of local models with the optimization capabilities of large cloud models.} Remarkably, our approach enables even weaker open-source models to achieve superior results compared to stronger proprietary baselines. 
In the optimization of critical path delay, our Qwen2.5-Instruct-7B + DeepSeek-V3 ensemble achieves a 50.85\% success rate, representing a 50\% improvement over standalone DeepSeek-V3 (33.90\%) and surpassing the proprietary GPT-4o baseline (45.76\%). For power optimization, the same combination reaches 66.67\% success, outperforming standalone DeepSeek-V3 (49.81\%) and approaching the performance of top-tier proprietary models like Gemini-2.5-Flash.
Similarly, our Qwen2.5-Coder-7B + DeepSeek-R1 combination demonstrates consistent improvements across both metrics. It achieves 44.07\% success in timing optimization (an 8.3\% improvement over DeepSeek-R1's 40.68\%) and 57.30\% success in power optimization (an 18.6\% improvement over DeepSeek-R1's 48.31\%).

\subsection{Analysis}
\paragraph{Effect of Model Types} We investigate the influence of model types by comparing models within the same family with different optimization objectives. As shown in Table~\ref{tab:main_ret}, our method consistently achieves performance improvements across multiple model combinations (Instruct+V3, Coder+R1, etc.), validating the broad applicability of our design principle framework.
Among local models, we find that Qwen2.5-Coder-7B does not show clear improvements over Qwen2.5-Instruct-7B, indicating that the task is relatively challenging and specialized code models may not excel, potentially due to the scarcity of high-quality Verilog code in the training corpus.
In contrast, when combined with our design principle framework, different cloud models exhibit distinct task-specific advantages. DeepSeek-V3 shows larger gains in timing performance optimization, while DeepSeek-R1 demonstrates greater improvements in power optimization. This indicates that different model architectures have complementary advantages under the guidance of our framework, allowing them to leverage their respective strengths for specific optimization objectives.

\paragraph{Effect of In-context Learning Example Number}
We investigate the impact of example number $K \in \{1, 2, 4, 8, 16, 32\}$ on optimization performance for different model combinations. 
Figure~\ref{fig:effect_varyK} reveals distinct, model-specific patterns that highlight the importance of tailored hyperparameter selection.
The DeepSeek-v3 + Qwen2.5-Instruct-7B configuration exhibits an inverted U-shaped relationship, peaking at $K$=4 for critical path delay (50.9\%). This suggests a trade-off between informativeness and noise: too few examples provide insufficient guidance, while excessive examples introduce conflicting signals. Power optimization shows less sensitivity to K, remaining stable across $K$=4-32.
In contrast, Qwen2.5-Instruct-7B + DeepSeek-R1 displays a U-shaped curve, achieving optimal performance at extremes ($K$=1: 54.2\% critical path, 61.1\% power; $K$=32: 50.9\% critical path, 62.5\% power) with degraded intermediate performance. We hypothesize this stems from DeepSeek-R1's reasoning-oriented architecture, which either benefits from minimal context to avoid interference or extensive context for comprehensive solution coverage. These findings underscore the necessity for model-specific ICL strategies, which we hope to explore in the future.

\begin{figure}[!t]
    \centering
    \includegraphics[width=0.9\linewidth]{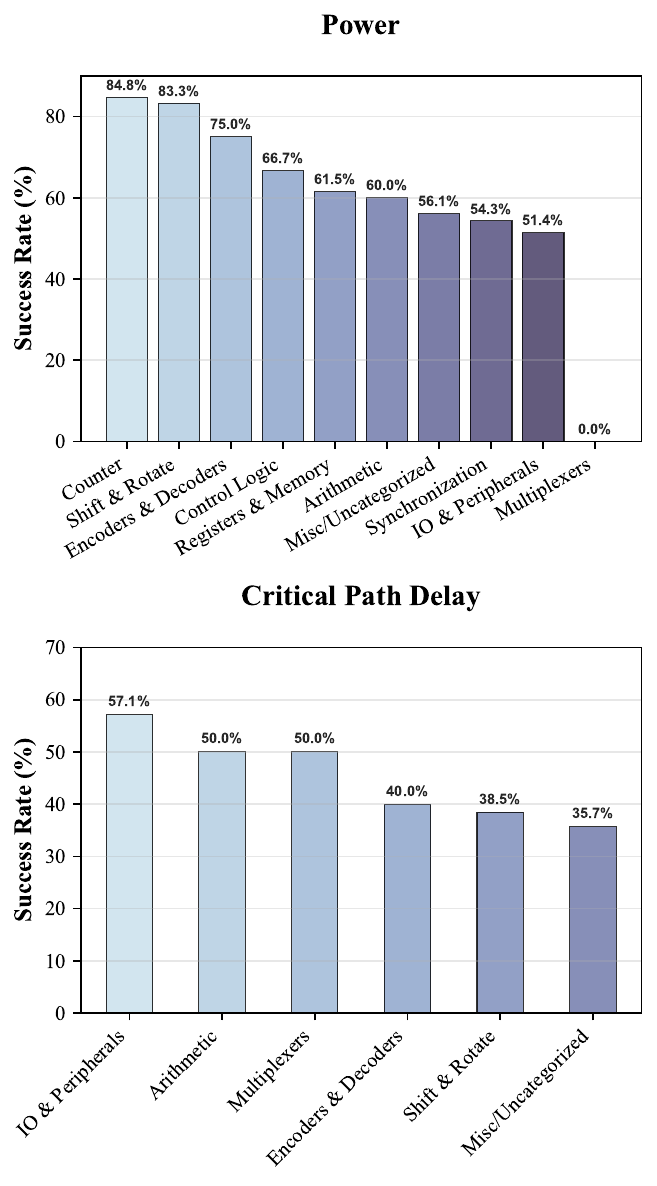}
    \caption{Optimization success rate across different module types (Qwen2.5-Instruct-7B + DeepSeek-V3).}
    \label{fig:llm_optimization_success_rates_sorted}
\end{figure}

\paragraph{Case Study}

We present six improved modules in Table~\ref{tab:case_study_ppa_improvement} with a representative case in Figure~\ref{fig:case_study}. Our method demonstrates that local models can extract effective design principles to guide cloud LLMs in achieving significant optimizations while protecting IP by never uploading actual RTL code.
Figure~\ref{fig:case_study} illustrates how our local model extracts key optimization principles that enable the cloud LLM to achieve a substantial 22.83\% critical path improvement (2.76 ns to 2.13 ns) by restructuring the calculator to pre-calculate arithmetic outcomes in parallel. This demonstrates the effectiveness of principle-based guidance over direct code optimization.
The extracted principles enable significant improvements across diverse optimization targets. For critical path delay, the method transforms a ripple-carry adder into a parallel-prefix architecture (3.12\% improvement) and collapses a multi-stage barrel shifter into single combinational logic (6.90\% improvement). For power optimization, more dramatic reductions are achieved: the display transmitter sees an 81.77\% power reduction (70.49 µW to 12.85 µW) by replacing FSM logic with streamlined shift operations, while the Schmitt trigger achieves 46.39\% reduction by eliminating redundant state storage. These results validate that our framework enables effective collaboration between local and cloud LLMs to perform architectural restructuring rather than superficial modifications.

\paragraph{Error Analysis}

The results in Figure~\ref{fig:llm_optimization_success_rates_sorted} reveal distinct patterns in LLM optimization capabilities. The model excels at local, pattern-based optimizations but struggles with tasks requiring global architectural changes or deep functional understanding.
For power optimization, the LLM achieves high success rates on regular structures like Counters (84.8\%) and Shift \& Rotate (83.3\%), where techniques like clock gating can be applied locally without altering core logic. However, performance degrades significantly for critical path delay optimization, with Shift \& Rotate success dropping to 38.5\%. This decline indicates the LLM cannot perform necessary global restructuring, such as converting serial shifters to parallel barrel shifters or introducing pipeline stages.
The moderate success on Synchronization modules (54.3\% for power) exposes a critical limitation: lack of functional understanding. While syntactically simple, synchronizers prevent metastability through careful design. The LLM likely attempts seemingly logical optimizations—such as removing ''redundant`` flip-flops—that violate the circuit's fundamental purpose, causing verification failures. This demonstrates that the LLM optimizes syntax without grasping critical hardware intent, revealing significant opportunities for developing better abstraction mechanisms that emphasize functional semantics.

\section{Conclusion}

This paper introduces a novel task of optimizing Verilog code for specific attributes while preserving IP data confidentiality, addressing a critical challenge in the semiconductor industry. We curate a comprehensive benchmark evaluating various LLMs and demonstrate that current approaches face a fundamental trade-off: large cloud models achieve superior optimization but require sensitive data sharing, while local models preserve privacy but exhibit limited performance.
Our proposed cloud-edge collaborative framework addresses this limitation by using local models to generate design principles that guide cloud models in optimization without exposing proprietary information. Experimental results show our method achieves up to 66.67\% power optimization success rate, outperforming even IP-leaking baselines. Error analysis reveals promising directions for future improvements in design principle generation and hint formulation strategies.
This work establishes a new research direction in LLM for EDA, enabling practical AI-assisted Verilog optimization in industrial settings.

\bibliography{aaai25}
\clearpage
\appendix 
\section{Dataset Sample}
We provide visualized Verilog codes for the barrel shifter designs and corresponding PPA metrics in Table~\ref{tab:barrel_shiter}.
\vspace{0.2in}

\begin{lstlisting}[style=verilog, caption=Verilog Implementation 1 - Multi-Stage Barrel Shifter]
module barrel_shifter_16bit (
    input [15:0] data,
    input [3:0] shift_amount,
    output reg [15:0] out
);

reg [15:0] stage1_out;
reg [15:0] stage2_out;

always @(*) begin
    stage1_out = (shift_amount[3]) ? data << 8 : data;
end

always @(*) begin
    stage2_out = (shift_amount[2]) ? stage1_out << 4 : stage1_out;
end

always @(*) begin
    out = (shift_amount[1]) ? stage2_out << 2 : stage2_out;
end

endmodule
\end{lstlisting}

\begin{lstlisting}[style=verilog, caption=Verilog Implementation 2 - Direct Barrel Shifter]
module barrel_shifter_16bit (
    input [15:0] data,
    input [3:0] shift_amount,
    output reg [15:0] out
);

    always @(*) begin
        if (shift_amount >= 0) begin
            out = data << shift_amount;
        end else begin
            out = data >> -shift_amount;
        end
    end

endmodule
\end{lstlisting}

\begin{table}[!htb]
\centering
\begin{tabular}{@{}lcc@{}}
\toprule
\textbf{Metric} & \textbf{Implementation 1} & \textbf{Implementation 2} \\
\midrule
\textbf{Area} & 34.902000 $\mu$m² & 44.856000 $\mu$m² \\
\textbf{Power} & 23.1250 $\mu$W & 29.0740 $\mu$W \\
\textbf{Clock Period} & 350.0000 ps & 390.0000 ps \\
\bottomrule
\end{tabular}
\caption{PPA comparison for barrel shifter module implementations.}
\label{tab:barrel_shiter}
\end{table}

\section{Prompt}

We provide the prompts used to query local LLMs to extract the design principles for  (i) low power consumption design (Table~\ref{tab:low_power}) and for (ii) low critical path delay (Table~\ref{tab:low_CPD}), according to the paired context examples.
The cloud LLMs are then prompted using the extracted design insights to improve the target attribute for a draft code, as shown in Table~\ref{tab:apply_insights}.

\begin{table}[t!]
\small
\centering
\begin{tcolorbox}[colback=lightgray, colframe=black, boxrule=1pt, arc=3pt]

You are an expert in Verilog design specializing in low-power optimization.

I will show you multiple pairs of Verilog implementations where one achieves 
significantly lower power consumption than the other. Your task is to analyze 
these examples and extract general design principles for low-power Verilog 
coding.

LEARNING CONTEXT - Power-Efficient vs Power-Inefficient Examples:

\textbf{\{context\_examples\}}

Based on these examples, provide general, high-level lessons or best practices 
for writing low-power Verilog code. Focus on:

1. \textcolor{highlightcolor}{Pattern Recognition}: What coding patterns consistently lead to better power efficiency?

2. \textcolor{highlightcolor}{Design Principles}: What general principles can be applied across different designs?

3. \textcolor{highlightcolor}{Common Pitfalls}: What practices should be avoided for power optimization?

4. \textcolor{highlightcolor}{Optimization Strategies}: What strategies work well for reducing power consumption?

Provide actionable, generalizable design principles that could be applied to 
the current design task. Do not reference specific code details from the 
examples, but instead extract the underlying principles that make 
implementations more power-efficient.

\end{tcolorbox}
\caption{Prompts for summarizing principles to reduce power consumption.}
\label{tab:low_power}
\end{table}

\begin{table}[tbh!]
\small
\centering
\begin{tcolorbox}[colback=lightgray, colframe=black, boxrule=1pt, arc=3pt]

You are an expert in Verilog design specializing in high-performance 
optimization.

I will show you multiple pairs of Verilog implementations where one achieves 
significantly lower critical path delay than the other. Your task is to 
analyze these examples and extract general design principles for 
high-performance Verilog coding.

LEARNING CONTEXT - Power-Efficient vs Power-Inefficient Examples:

\textbf{\{context\_examples\}}

Based on these examples, provide general, high-level lessons or best practices 
for writing low-power Verilog code. Focus on:

1. \textcolor{highlightcolor}{Pattern Recognition}: What coding patterns consistently lead to better power efficiency?

2. \textcolor{highlightcolor}{Design Principles}: What general principles can be applied across different designs?

3. \textcolor{highlightcolor}{Common Pitfalls}: What practices should be avoided for power optimization?

4. \textcolor{highlightcolor}{Optimization Strategies}: What strategies work well for reducing power consumption?

Provide actionable, generalizable design principles that could be applied to 
the current design task. Do not reference specific code details from the 
examples, but instead extract the underlying principles that make 
implementations more power-efficient.

\end{tcolorbox}
\caption{Prompts for summarizing principles to reduce critical delay path.}
\label{tab:low_CPD}
\end{table}

\begin{table}[tbh!]
\small
\centering
\begin{tcolorbox}
Here is a general lesson for writing \textbf{\{target\}} Verilog:

\textbf{\{design principle\}}

Please optimize the following Verilog code for \{target\}, following the above 
lesson. 
Do not copy any code from other sources, only improve the given 
code.

Verilog code:

\textasciigrave\textasciigrave\textasciigrave verilog

\textbf{\{verilog to optimize\}}

\textasciigrave\textasciigrave\textasciigrave

\end{tcolorbox}
\caption{Prompts for cloud LLMs to apply the summarized design principles to improve the Verilog code regarding the target attribute (e.g., power consumption or critical path delay).}
\label{tab:apply_insights}
\end{table}

\end{document}